\RequirePackage{fix-cm}
\documentclass[smallextended]{svjour3}       
\smartqed  
\usepackage{graphicx}
\usepackage{cite}
\usepackage{amsmath,amssymb,amsfonts}
\usepackage{algorithmic}
\usepackage{graphicx}
\usepackage{textcomp}
\def\BibTeX{{\rm B\kern-.05em{\sc i\kern-.025em b}\kern-.08em
    T\kern-.1667em\lower.7ex\hbox{E}\kern-.125emX}}
\RequirePackage[table]{xcolor}
\usepackage{booktabs}
\usepackage[hidelinks]{hyperref} 
\usepackage{natbib}
\usepackage[misc]{ifsym}

\usepackage{nicefrac}
\usepackage{svg}
\usepackage{pgfplots}
\pgfplotsset{compat=1.12}
\usepackage{rotating}
\newcommand*\rot{\rotatebox{90}}
\usepackage{mdframed}
\mdfdefinestyle{summarybox}{%
  rightline=true,
  innerleftmargin=10,
  innerrightmargin=10,
  linewidth=2pt,
  topline=false,
  bottomline=false,
  skipabove=\topsep,
  skipbelow=\topsep,
  frametitlefont=\normalfont\sc
}
\usepackage{array,hhline}
\usepackage{pgf}

\newcommand*{\gray}{gray}

\newcommand*{\colorme}[1]{%
    \pgfmathparse{#1<62?1:0}%
    \ifnum\pgfmathresult=0\relax\color{white}\fi
    \pgfmathparse{(127-#1)/127}
    \expandafter\cellcolor\expandafter[\expandafter\gray\expandafter]\expandafter{\pgfmathresult}%
    #1%
}

\newlength{\tabwidth}
\begin{document}

\title{Automatically Categorising GitHub Repositories by \\ Application Domain}

\author{Francisco Zanartu \and Christoph Treude~\Letter \and Bruno Cartaxo \and Hudson Silva Borges \and Pedro Moura \and Markus Wagner \and Gustavo Pinto}

\institute{
Francisco Zanartu \at The University of Adelaide, Australia\\ \email{francisco.zanartu@adelaide.edu.au} \and
\Letter~Corresponding author - Christoph Treude \at The University of Melbourne, Australia\\ \email{christoph.treude@unimelb.edu.au} \and
Bruno Cartaxo \at Federal Institute of Pernambuco, Brazil\\ \email{email@brunocartaxo.com} \and
Hudson Silva Borges \at Federal University of Mato Grosso do Sul, Brazil\\ \email{hsborges@facom.ufms.br} \and
Pedro Moura \at Federal Institute of Pernambuco, Brazil\\ \email{ppomoura@gmail.com} \and
Markus Wagner \at The University of Adelaide, Australia\\ \email{markus.wagner@adelaide.edu.au} \and
Gustavo Pinto \at Federal University of Pará, Brazil \& Zup Innovation, Brazil\\ \email{gpinto@ufpa.br}
}

\maketitle

\begin{abstract}
GitHub is the largest host of open source software on the Internet. This large, freely accessible database has attracted the attention of practitioners and researchers alike. But as GitHub's growth continues, it is becoming increasingly hard to navigate the plethora of repositories which span a wide range of domains. Past work has shown that taking the application domain into account is crucial for tasks such as predicting the popularity of a repository and reasoning about project quality. In this work, we build on a previously annotated dataset of 5,000 GitHub repositories to design an automated classifier for categorising repositories by their application domain.  The classifier uses state-of-the-art natural language processing techniques and machine learning to learn from multiple data sources and catalogue repositories according to five application domains. We contribute with (1) an automated classifier that can assign popular repositories to each application domain with at least 70\% precision, (2) an investigation of the approach's performance on less popular repositories, and (3) a practical application of this approach to answer how the adoption of software engineering practices differs across application domains. Our work aims to help the GitHub community identify repositories of interest and opens promising avenues for future work investigating differences between repositories from different application domains.
\end{abstract}

\begin{sloppy}

\section{Introduction and Motivation}

GitHub is the largest host of open source code on the Internet~\citep{kim2021sequential} and has piqued the interest of practitioners and researchers, many of whom have struggled to bring structure to the plethora of repositories available on the platform. For example, there are limited means available to separate repositories containing engineered software projects from other repositories, such as personal projects or those that use GitHub for free cloud storage~\citep{kalliamvakou2014promises, munaiah2017curating}. To make it easier for users to identify relevant repositories for their wide variety of use cases, GitHub has been adding features to its service, such as README files, topics tags, and showcases (where contributors describe, add keywords, and label their repository). However, these features are insufficient for many use cases.

For example, while achieving generalizability of the results is the primary objective of many empirical papers, modern computing research is largely application domain independent~\citep{capiluppi2020using}. Application domains are the sections of reality for which a software system is designed. Their importance relies on their serving as the starting point for actual state analysis and usually includes domain-specific language, meaning that developers in this domain think about their project in a specific way, with particular terms and concepts~\citep{Zullighoven2005}.  Application domains are not a feature currently implemented by GitHub to catalogue repositories. Previous work has found that repository quality indicators, such as object-oriented metrics, can be ``extremely sensitive to application domains''~\citep{capiluppi2019relevance}, and that the application domain is an important factor in predicting repository popularity~\citep{borges2016understanding}.  Furthermore, since documentation of GitHub repositories is often incomplete~\citep{prana2019categorizing}, information about the application domain of a repository can be crucial to gain a high-level understanding of its content and purpose. \cite{borges2016understanding} manually annotated 5,000 GitHub repositories according to their application domain, distinguishing repositories into \textsc{Application Software}, \textsc{System Software}, \textsc{Web Libraries and Frameworks}, \textsc{Non-Web Libraries and Frameworks}, \textsc{Software Tools}, and \textsc{Documentation}. In this paper, we use this previously annotated dataset to develop a machine learning solution to automatically classify GitHub repositories by their application domain. In particular, our contributions are the following:

\begin{itemize}
    \item an automated classifier which can assign popular GitHub repositories to an application domain with at least 70\% precision for each of the application domains;
    \item an investigation of the performance of the approach on a newly annotated dataset of less popular repositories; and
    \item a study that puts our classifier into practice by investigating how recent and popular repositories, associated with different application domains, differ in their adoption of software engineering practices, such as automation, refactoring, and code ownership.
\end{itemize}

Our work aims to help researchers and practitioners identify relevant repositories for their use cases, offering a trade-off between accuracy and volume to many who have manually annotated repositories using \cite{borges2016understanding}'s classification~\citep{bi2020empirical, bi2021first, hayashi2019impacts, rehman2022newcomer, hata2022github, de2022impacts, de2020splicing, tamburri2019discovering, coelho2017modern} or intend to use it in future work~\citep{wu2020dockerfile, wu2020exploring, wu2020characterizing, bayati2019time, zhang2020github, de2022impacts}. For example, \cite{bi2020empirical} manually annotated 1,000 repositories in terms of their application domain using the Borges et al.~taxonomy to investigate release note production and usage of release notes in practice, and \cite{hayashi2019impacts} manually annotated 969 repositories in terms of their application domain to investigate the impacts of daylight saving time on software development. Both of these had to invest substantial manual effort in the annotation task that our work automates. In addition, our investigation of differences in the adoption of software engineering practises across application domains opens up promising avenues for future work, aiding in the achievement of generalisable results for the selected software engineering domain. All data and scripts are available on Zenodo.\footnote{\url{https://doi.org/10.5281/zenodo.6423599}}

The remainder of this paper is structured as follows. We introduce the dataset used as the starting point of this work in Section~\ref{sec:background} and our research questions in Section~\ref{sec:rqs}. We outline our methods for data collection, data pre-processing, and data analysis in Section~\ref{sec:methodology} and also provide a characterisation of the data in the same section. Sections~\ref{sec:rq1} through~\ref{sec:domaindiffs} provide answers to our research questions. Section~\ref{sec:discussion} discusses our results and Section~\ref{sec:threats} the threats to the validity of this work, and Section~\ref{sec:related} places our work in the context of existing literature. Section~\ref{sec:conclusion} concludes this paper and describes future work.

\section{Application Domains}
\label{sec:background}

We use the existing dataset curated by \cite{borges2016understanding} as the starting point for our work to build an automated classifier for categorising GitHub repositories. To help repository owners and clients understand how their repositories are performing in a competitive open-source development market, the authors developed a multiple linear regression classifier to predict the number of stars of GitHub repositories. One of the features used in their prediction model is the repositories' application domain, which the authors determined by manually annotating 5,000 repositories with their respective application domain. Figure~\ref{fig:domaindistribution} summarises the distribution of the application domains in this dataset.

\begin{figure}
\centering
\begin{tikzpicture}
  \begin{axis}[
    xbar,
    x = 0.045,
    y = 15,
    y axis line style = { opacity = 0 },
    ytick = data,
    axis x line = none,
    tickwidth = 0pt,
    symbolic y coords = {System Software, Documentation, Application Software, Software Tools, Non-Web Libraries and Frameworks, Web Libraries and Frameworks},
    nodes near coords,
  ]
  \addplot[color=black, fill=black] coordinates {(179,System Software) (427,Documentation) (428,Application Software) (963,Software Tools) (1429,Non-Web Libraries and Frameworks) (1522,Web Libraries and Frameworks)};
  \end{axis}
\end{tikzpicture}
\caption{Distribution of application domains in the original dataset}
\label{fig:domaindistribution}
\end{figure}
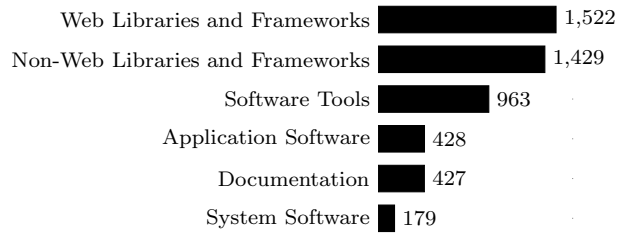

The dataset contains eight features for each repository in the sample. In addition to stars and the application domain, it includes repository names, the number of forks, the main programming language, a short description, URL, and the growth pattern. The six application domains are explained by the authors as follows~\citep{borges2016understanding}:

\begin{itemize}
    \item \textsc{Application Software}: systems that provide functionalities to end-users, like browsers and text editors (e.g., \textsc{wordpress/wordpress} and \textsc{adobe/brackets}).
    \item \textsc{System Software}: systems that provide services and infrastructure to other systems, like operating systems, middleware, servers, and databases (e.g., \textsc{torvalds/linux} and \textsc{mongodb/mongo}).
    \item \textsc{Web Libraries and Frameworks} (e.g., \textsc{twbs/bootstrap} and \textsc{angular/angular.js}).
    \item \textsc{Non-Web Libraries and Frameworks} (e.g., \textsc{google/guava} and \textsc{facebook/fresco}).
    \item \textsc{Software Tools}: systems that support software development tasks, like IDEs, package managers, and compilers (e.g., \textsc{homebrew/homebrew} and \textsc{git/git}).
    \item \textsc{Documentation}: repositories with documentation, tutorials, source code examples, etc. (e.g., \textsc{iluwatar/java-design-patterns}).
\end{itemize}

We use this annotated dataset for our work on developing an automated classifier for the application domain of a GitHub repository and as the starting point of our investigation of differences among the application domains. The Borges et al.~dataset is publicly available on Zenodo.\footnote{\url{https://doi.org/10.5281/zenodo.804474}}

\section{Research Questions}
\label{sec:rqs}

We ask two research questions (RQ) to guide this work, focusing on automatically classifying GitHub repositories and evaluating the performance and potential applications of the classifier.

\begin{description}
\item[RQ1] How accurately can we automatically classify GitHub repositories?
\end{description}

We investigate the classifier performance on the dataset annotated by \cite{borges2016understanding} and on our newly annotated dataset of less popular repositories, since the original dataset considered only the most popular. In addition, we investigate to what extent the classifier performance improves if we merge closely related classes that cause confusion for the classifier and the human annotators.

\begin{description}
\item[RQ1.1] What is the impact of adding different data sources on the classifier performance?
\item[RQ1.2] How accurately can state-of-the-art machine learning techniques automatically classify popular GitHub repositories?
\item[RQ1.3] What is the impact of merging closely related classes of GitHub repositories?
\item[RQ1.4] What is the importance of different features for the classification?
\item[RQ1.5] How accurately can state-of-the-art machine learning techniques automatically classify less popular GitHub repositories?
\end{description}

The results of RQ1 provide an understanding of the features that the model considers critical to obtaining the results. These findings should be taken into account when explaining results in repositories that may contain incomplete documentation, lowering the classifier's performance.

RQ2 investigates differences between repositories from different application domains from the point of view of practical applications of the classifier:

\begin{description}
\item[RQ2] How does the adoption of software engineering practices differ between application domains?
\end{description}

The research question focuses on automation, refactoring, and code ownership, which are some of the most widely adopted software engineering practices. 
We pose distinct RQs for each. Ultimately, RQ2 aims to demonstrate how the classification of GitHub repositories could be used to answer additional research questions in the future.

\begin{description}
\item[RQ2.1] How does the adoption of GitHub Actions differ between application domains?
\item[RQ2.2] How does the extent of refactoring differ between application domains?
\item[RQ2.3] How does code ownership differ between application domains?
\end{description}

\section{Methodology}
\label{sec:methodology}

In this section, we describe our procedure for data collection and characterise the collected data. We also outline our methods for data pre-processing and data analysis.

\subsection{Data collection}

We use the \cite{borges2016understanding} data set as the basis of our work. We then used the GitHub API\footnote{\url{https://docs.github.com/en/rest}} via the PyGitHub library\footnote{\url{https://github.com/PyGithub/PyGithub}}
to enrich and expand this dataset. 

Our choice of features to extract for each repository was inspired by previous work on processing GitHub repositories, such as \cite{prana2019categorizing} and \cite{zhang2017detecting}.

Table~\ref{tab:textfeatures} provides a description of the text and categorical features that we obtained from GitHub to construct our dataset. README files, descriptions, and labels are further processed using natural language processing techniques, which we will describe below.

\begin{table}
\centering
\caption{Textual and categorical features of GitHub repositories}
\label{tab:textfeatures}
\begin{tabular}{lp{5.6cm}}
\toprule
Feature & Description \\
\midrule
Description     & A short text that should briefly tell the public what it is contained in the repository. \\
README File     & The text file that GitHub shows when someone navigates to a repository. It commonly contains a set of helpful information about the repository and usage information.  \\
Topics          & Topics are labels that let users explore repositories by those labels, thus creating subject-based connections between GitHub repositories. \\
Licence         & A licence is a fundamental part of the definition of open-source software. \\
Prog.~Languages & GitHub provides repository statistics about the programming languages used in files within a repository. \\
Labels          & Labels are intended to help users categorise issues, pull requests, and discussions. There is also a list of nine default labels that create a standard workflow in a repository, such as a bug, documentation, and duplicate. \\
Contributors    & Contributors are the usernames of non-anonymous users who contribute to GitHub repositories. \\
Sub-folders/files     & Several files and folders could make up a repository; this feature lists all the sub-folder and file names in the root folder. \\
\bottomrule
\end{tabular}
\end{table}

We also consider numeric features such as releases, stars, and forks, summarised in Table~\ref{tab:numfeatures}. 

\begin{table}
\centering
\caption{Numerical features of GitHub repositories}
\label{tab:numfeatures}
\begin{tabular}{lp{5.6cm}}
\toprule
Feature & Description \\
\midrule
Releases       & Releases are tags that mark a specific point in the repository history. This feature is the count of those particular points. \\
Stars          & A star is a bookmark or display of appreciation for a repository.\\
Forks          & Forks are individual copies of another user's repository. Forking allows making changes to the source code without affecting the original root repository.   \\
\bottomrule
\end{tabular}
\end{table}

\subsection{Data characterisation}

After completing the data collection described in the previous section, we conducted a manual inspection to remove repositories from the dataset that are no longer maintained, as indicated by descriptions or README files where the original content was replaced with messages such as ``This repository is no longer maintained, has been removed, or is deprecated.'' This step did not substantially affect the distribution of application domains, but was necessary to avoid misleading values in the training data, such as empty repositories labelled with a particular application domain. As a result of this step, the sample was reduced to 4,948, as shown in Table~\ref{tab:reduction}.

\begin{table}
\centering
\caption{Removal of repositories no longer available}
\label{tab:reduction}
\begin{tabular}{lrrrr}
\toprule
Application Domain                           & Samples & \% & Previous & Diff. \\
\midrule
Web Libs \& Frameworks          & 1,522           & 31       & 1,535                    & 13         \\
Non-Web Libs \& Frameworks         & 1,429           & 29       & 1,439                    & 10         \\
Software Tools                     & 963             & 19       & 972                      & 9          \\
Application Software               & 428             & 9       & 437                      & 9          \\
Documentation                      & 427             & 9       & 433                      & 6          \\
System Software                    & 179             & 4       & 184                      & 5          \\
\midrule
Total                             & 4,948           & 100       & 5,000                    & 52        \\
\bottomrule
\end{tabular}
\end{table}

Figure~\ref{fig:domaindistribution} shows that the dataset is imbalanced. The most frequent application domain, \textsc{Web Libraries and Frameworks}, as well as \textsc{Non-Web Libraries and Frameworks}, account for 60\% of the dataset. The least frequent application domain, \textsc{System Software}, accounts for only 4\% of the dataset. 

To further characterise the data, Table~\ref{tab:missing} shows how many values are missing because the owners of the repository did not provide this information. The table supports \cite{zhang2019higitclass}'s discovery that the majority of GitHub repositories lack topic tags.

\begin{table}
\centering
\caption{Missing values per feature}
\label{tab:missing}
\begin{tabular}{lrrr}
\toprule
Data source           & Values & Missing & (\%) \\
\midrule
Description           & 4,914         & 34             & 0.7        \\
README File           & 4,948         & 0              & 0          \\
Topics                & 2,335         & 2,613          & 52.8       \\
Licence               & 4,301         & 647            & 13.1       \\
Programming Languages & 4,898         & 50             & 1          \\
Labels                & 4,720         & 228            & 4.6        \\
Contributors          & 4,942         & 6              & 0.1        \\
Sub-folders/files     & 4,948         & 0              & 0          \\
Releases              & 4,948         & 0              & 0          \\
Stars                 & 4,948         & 0              & 0          \\
Forks                 & 4,948         & 0              & 0          \\
\bottomrule
\end{tabular}
\end{table}


Most of the GitHub repositories in our dataset use a licence to share their code.\footnote{Note that we only count repositories that have GitHub's licence feature activated. Some repositories declare their licence informally inside of a README file.} They use five programming languages on average, contain 22 sub-folders and files in their root folder, and have 95 non-anonymous contributors per repository. Table~\ref{tab:textfeaturevalues} summarises this information and displays the number of unique words 
provided by each data source, with ``Sub-folders/files'' and ``Contributors'' being high-dimensional data sources. Note that each word in each of the textual sources potentially becomes a feature used in the classifier, i.e., we employed one-hot encoding.

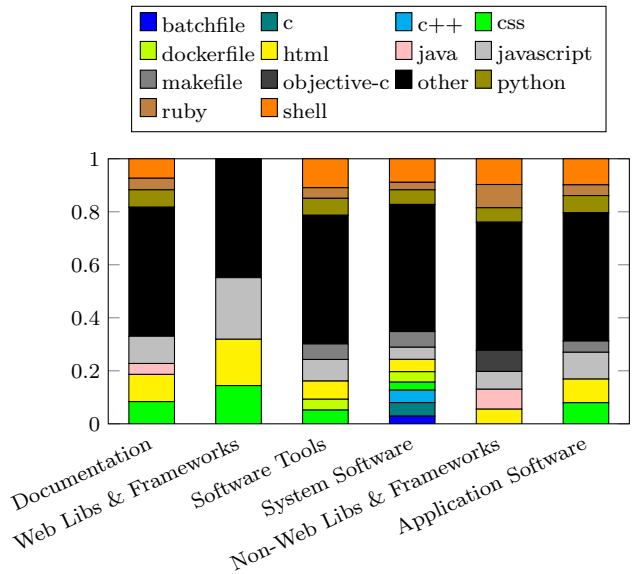
\begin{figure}
    \centering\small
    \begin{tikzpicture}
    \begin{axis}[
    ybar stacked, ymin=0, ymax=1, y=100,
    x tick label style={rotate=25,anchor=north east},
    bar width=6mm,
    symbolic x coords={Documentation, Web Libs \& Frameworks, Software Tools, System Software, Non-Web Libs \& Frameworks, Application Software},
    xtick=data,
    legend style={at={(0.5,1.1)},anchor=south,
    legend cell align={left},
    legend columns=4}
    ]
    \addplot [fill=blue] coordinates {
        (Documentation,0)
        (Web Libs \& Frameworks,0)
        (Software Tools,0)
        (System Software,0.02900232)
        (Non-Web Libs \& Frameworks,0)
        (Application Software,0)
    };
    \addplot [fill=teal] coordinates {
        (Documentation,0)
        (Web Libs \& Frameworks,0)
        (Software Tools,0)
        (System Software,0.051044084)
        (Non-Web Libs \& Frameworks,0)
        (Application Software,0)
    };
    \addplot [fill=cyan] coordinates {
        (Documentation,0)
        (Web Libs \& Frameworks,0)
        (Software Tools,0)
        (System Software,0.046983759)
        (Non-Web Libs \& Frameworks,0)
        (Application Software,0)
    };
    \addplot [fill=green] coordinates {
        (Documentation,0.083632019)
        (Web Libs \& Frameworks,0.144128788)
        (Software Tools,0.051808407)
        (System Software,0.030162413)
        (Non-Web Libs \& Frameworks,0)
        (Application Software,0.07960199)
    };
    \addplot [fill=lime] coordinates {
        (Documentation,0)
        (Web Libs \& Frameworks,0)
        (Software Tools,0.040860215)
        (System Software,0.039443155)
        (Non-Web Libs \& Frameworks,0)
        (Application Software,0)
    };
    \addplot [fill=yellow] coordinates {
        (Documentation,0.102747909)
        (Web Libs \& Frameworks,0.174810606)
        (Software Tools,0.068621701)
        (System Software,0.046403712)
        (Non-Web Libs \& Frameworks,0.055218216)
        (Application Software,0.089137645)
    };
    \addplot [fill=pink] coordinates {
        (Documentation,0.041218638)
        (Web Libs \& Frameworks,0)
        (Software Tools,0)
        (System Software,0)
        (Non-Web Libs \& Frameworks,0.075142315)
        (Application Software,0)
    };
    \addplot [fill=lightgray] coordinates {
        (Documentation,0.102747909)
        (Web Libs \& Frameworks,0.232386364)
        (Software Tools,0.081524927)
        (System Software,0.045823666)
        (Non-Web Libs \& Frameworks,0.067172676)
        (Application Software,0.101160862)
    };
    \addplot [fill=gray] coordinates {
        (Documentation,0)
        (Web Libs \& Frameworks,0)
        (Software Tools,0.058455523)
        (System Software,0.059164733)
        (Non-Web Libs \& Frameworks,0)
        (Application Software,0.042288557)
    };
    \addplot [fill=darkgray] coordinates {
        (Documentation,0)
        (Web Libs \& Frameworks,0)
        (Software Tools,0)
        (System Software,0)
        (Non-Web Libs \& Frameworks,0.079696395)
        (Application Software,0)
    };
    \addplot [fill=black] coordinates {
        (Documentation,0.486857826)
        (Web Libs \& Frameworks,0.448674242)
        (Software Tools,0.485630499)
        (System Software,0.479698376)
        (Non-Web Libs \& Frameworks,0.483491461)
        (Application Software,0.484245439)
    };
    \addplot [fill=olive] coordinates {
        (Documentation,0.065710872)
        (Web Libs \& Frameworks,0)
        (Software Tools,0.063929619)
        (System Software,0.055104408)
        (Non-Web Libs \& Frameworks,0.054459203)
        (Application Software,0.064676617)
    };
    \addplot [fill=brown] coordinates {
        (Documentation,0.043608124)
        (Web Libs \& Frameworks,0)
        (Software Tools,0.040078201)
        (System Software,0.028422274)
        (Non-Web Libs \& Frameworks,0.087666034)
        (Application Software,0.040630182)
    };
    \addplot [fill=orange] coordinates {
        (Documentation,0.073476703)
        (Web Libs \& Frameworks,0)
        (Software Tools,0.109090909)
        (System Software,0.0887471)
        (Non-Web Libs \& Frameworks,0.0971537)
        (Application Software,0.098258706)
    };
    \legend{batchfile,c,c++,css,dockerfile,html,java,javascript,makefile,objective-c,other,python,ruby,shell}
    \end{axis}
 \end{tikzpicture}
    \caption{Most frequent programming languages per application domain}
    \label{fig:languagesperdomain}
\end{figure}

The programming languages used might also be an indicator of the application domain of a repository. For example, \textsc{Web Libraries and Frameworks} account for JavaScript, HTML, and CSS as the languages that are the most used. \textsc{System Software}, on the other hand, comes in a variety of languages (e.g., Shell, Ruby, C, C++). However, the other application domains do not differ substantially. Figure~\ref{fig:languagesperdomain} presents this information in visual form.

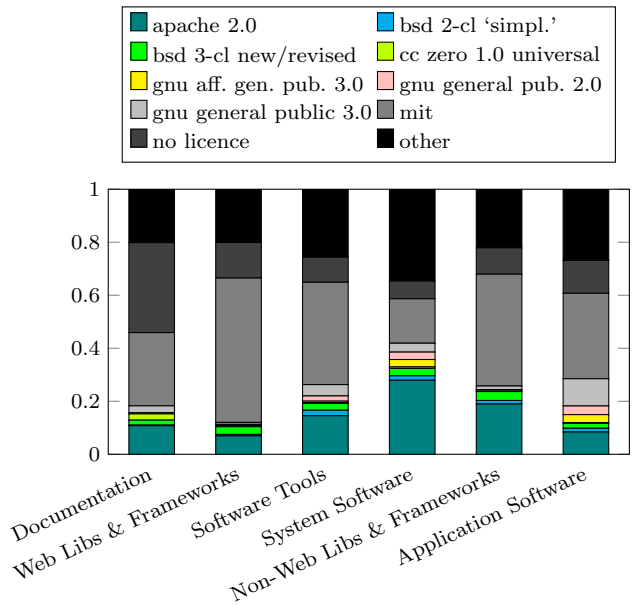
\begin{figure}
    \centering\small
    \begin{tikzpicture}
    \begin{axis}[
    ybar stacked, ymin=0, ymax=1, y=100,
    x tick label style={rotate=25,anchor=north east},
    bar width=6mm,
    symbolic x coords={Documentation, Web Libs \& Frameworks, Software Tools, System Software, Non-Web Libs \& Frameworks, Application Software},
    xtick=data,
    legend style={at={(0.5,1.1)},anchor=south,
    legend cell align={left},
    legend columns=2}
    ]
    \addplot [fill=teal] coordinates {
        (Application Software,0.08411215)
        (Documentation,0.107728337)
        (Non-Web Libs \& Frameworks,0.189643107)
        (Software Tools,0.145379024)
        (System Software,0.279329609)
        (Web Libs \& Frameworks,0.069645204)
    };
    \addplot [fill=cyan] coordinates {
        (Application Software,0.014018692)
        (Documentation,0.00234192)
        (Non-Web Libs \& Frameworks,0.013296011)
        (Software Tools,0.020768432)
        (System Software,0.016759777)
        (Web Libs \& Frameworks,0.005256242)
    };
    \addplot [fill=green] coordinates {
        (Application Software,0.018691589)
        (Documentation,0.018735363)
        (Non-Web Libs \& Frameworks,0.033589923)
        (Software Tools,0.02596054)
        (System Software,0.027932961)
        (Web Libs \& Frameworks,0.02890933)
    };
    \addplot [fill=lime] coordinates {
        (Application Software,0.002336449)
        (Documentation,0.023419204)
        (Non-Web Libs \& Frameworks,0)
        (Software Tools,0.002076843)
        (System Software,0.005586592)
        (Web Libs \& Frameworks,0.003285151)
    };
    \addplot [fill=yellow] coordinates {
        (Application Software,0.030373832)
        (Documentation,0)
        (Non-Web Libs \& Frameworks,0.00209937)
        (Software Tools,0.00623053)
        (System Software,0.027932961)
        (Web Libs \& Frameworks,0.002628121)
    };
    \addplot [fill=pink] coordinates {
        (Application Software,0.03271028)
        (Documentation,0.004683841)
        (Non-Web Libs \& Frameworks,0.005598321)
        (Software Tools,0.01973001)
        (System Software,0.027932961)
        (Web Libs \& Frameworks,0.003285151)
    };
    \addplot [fill=lightgray] coordinates {
        (Application Software,0.102803738)
        (Documentation,0.025761124)
        (Non-Web Libs \& Frameworks,0.013296011)
        (Software Tools,0.042575286)
        (System Software,0.033519553)
        (Web Libs \& Frameworks,0.007227332)
    };
    \addplot [fill=gray] coordinates {
        (Application Software,0.322429907)
        (Documentation,0.276346604)
        (Non-Web Libs \& Frameworks,0.421973408)
        (Software Tools,0.386292835)
        (System Software,0.167597765)
        (Web Libs \& Frameworks,0.544678055)
    };
    \addplot [fill=darkgray] coordinates {
        (Application Software,0.123831776)
        (Documentation,0.339578454)
        (Non-Web Libs \& Frameworks,0.099370189)
        (Software Tools,0.094496366)
        (System Software,0.067039106)
        (Web Libs \& Frameworks,0.134034166)
    };
    \addplot [fill=black] coordinates {
        (Application Software,0.268691589)
        (Documentation,0.201405152)
        (Non-Web Libs \& Frameworks,0.22113366)
        (Software Tools,0.256490135)
        (System Software,0.346368715)
        (Web Libs \& Frameworks,0.201051248)
    };
    \legend{apache 2.0,bsd 2-cl~`simpl.',bsd 3-cl~new/revised,cc zero 1.0 universal,gnu aff.~gen.~pub.~3.0,gnu general pub.~2.0,gnu general public 3.0,mit,no licence,other}
    \end{axis}
 \end{tikzpicture}
    \caption{Most frequent licences per application domain}
    \label{fig:licences}
\end{figure}

Figure~\ref{fig:licences} shows that the MIT licence is the most frequently adopted licence in the dataset, with the exception of the application domain \textsc{System Software}. \textsc{Documentation} repositories frequently lack a licence. 

\begin{table}
\centering
\caption{Values of textual features}
\label{tab:textfeaturevalues}
\begin{tabular}{lr@{\hspace{1em}}r@{\hspace{1em}}r@{\hspace{1em}}r@{\hspace{1em}}r}
\toprule
                                 & \rot{Languages} & \rot{Licence} & \rot{Topics} & \rot{Sub-folders/files} & \rot{Contributors} \\
\midrule
Application Software             & 6         & 1       & 4      & 21       & 87           \\
Documentation                    & 4         & 1       & 3      & 23       & 76           \\
Non-Web Libs \& Frameworks       & 4         & 1       & 3      & 18       & 72           \\
Software Tools                   & 5         & 1       & 3      & 22       & 105          \\
System Software                  & 10        & 1       & 4      & 29       & 147          \\
Web Libs \& Frameworks           & 3         & 1       & 3      & 18       & 83           \\
\midrule
Average number of features       & 5         & 1       & 3      & 22       & 95           \\
Unique features                  & 336       & 24      & 6,036  & 27,752   & 239,364      \\
\bottomrule
\end{tabular}
\end{table}

Table~\ref{tab:textfeaturevalues} shows that \textsc{Software Tools} and \textsc{System Software} are the application domains with the most non-anonymous contributors. The box plot in Figure~\ref{fig:boxplot} shows the number of non-anonymous contributors per application domain, with a pronounced presence of outliers. Only \textsc{System Software} appears to be free of outliers, which is likely due to the small sample size (4\% of the dataset). Contributors are a critical component of collaborative open-source software development. While the majority of non-anonymous contributors have made few contributions, others collaborate on multiple repositories. These contributors, in particular, are of interest for this research because we might be able to establish that a particular contributor is more likely to contribute to repositories of a particular application domain.

With this data characterisation in mind, we next identify suitable pre-processing techniques for each feature or group of features in such a way that could be interpretable for a machine learning classifier. At first sight, we can identify text data from README files, repository description, and the labels used to classify issues, pull requests, and discussions. Contributors, programming languages, topic keywords, sub-folders/files names, and type of licence can be interpreted as categorical data for each repository, and finally, as numerical data, we will use the number of forks, stars, and releases. Hence, the challenge for our next step will be how to pre-process and combine all these features into an interpretable dataset for our classifier.

\begin{figure}
  \centering
  \includegraphics[width=\linewidth]{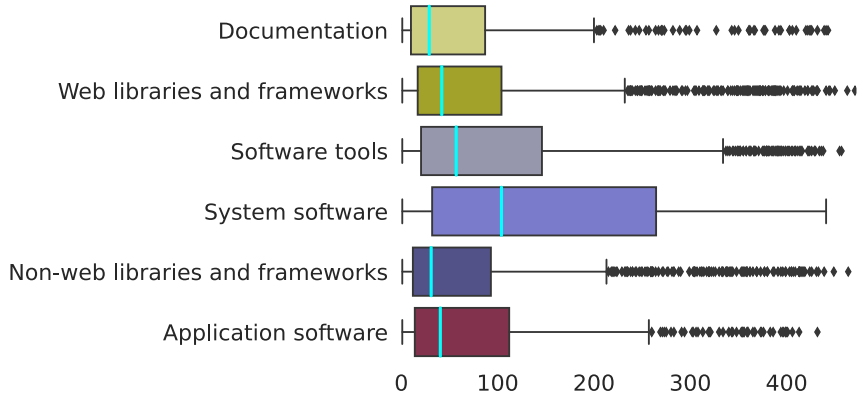}
  \caption{Non-anonymous contributors per application domain}
  \label{fig:boxplot}
\end{figure}

\subsection{Data pre-processing}
\label{sec:cleaning}

Textual data, i.e., README files, repository descriptions, and labels, require the application of natural language processing (NLP) techniques. There are several techniques for pre-processing and analysing natural language data, such as Bag-of-words, TF-IDF, or Word2Vec, as well as transformer models, which have achieved state-of-the-art performance in a wide range of NLP tasks~\citep{DBLP:journals/corr/VaswaniSPUJGKP17, DBLP:journals/corr/abs-1810-04805}. The Sentence Transformers Library is a low-cost option for obtaining the embedding representation of our text data that is not big enough to train a deep learning model~\citep{reimers-2019-sentence-bert}.

Before pre-processing text data, the README files are cleaned of HTML and Markdown tags as well as duplicated white spaces. Then, the embedding representation is obtained using the pre-trained model `all-mpnet-base-v2' which is an all-round model tuned from many use cases and exhibits the best average performance on a diverse benchmark dataset. This model returns a 768-dimensional vector for each sample.

We note that categorical data, such as topic labels assigned by repository maintainers, is mostly sparse due to the employed one-hot encoding, thus contradicting our objective of finding recurrent features that correlate with our application domains. Therefore, we perform feature selection, which has the added benefit of decreasing the chances of overfitting and leads to simpler models that generalise better~\citep{muller2016introduction}. According to the Scikit-learn library~\citep{pedregosa2011scikit}, model-based feature selection keeps only the most important features by studying not only univariate statistically significant relationships between each feature and label, but also the combination of all features with an importance measure greater than the provided threshold. We chose LinearSVM with C = 0.01 as the threshold to reduce sparsity in this work.

For contributors, we only used the 50 most prolific ones from each application domain as input to our model. Unsurprisingly, the distribution of contributions to repositories in different application domains follows a power-law distribution, with the vast majority of developers only contributing to a single repository and thus not providing us with useful information to learn from for classifying other repositories.

\subsection{Data analysis}
\label{sec:analysis}

To find an adequate solution for RQ1, we needed to address the problem of combined algorithm selection and hyperparameter tuning (CASH)~\citep{thornton2013auto}. The solution to the CASH problem intends to address two problems for automated machine learning (AutoML) methods: There is no unique machine learning algorithm that performs well on every dataset, and some machine learning algorithms need hyper-parameter optimisation to achieve good results~\citep{feurer2019auto}.

AutoML libraries usually require multiple trials of different configurations or long training times to find accurate models and outperform a tuned random forest baseline benchmark~\citep{gijsbers2019open}. In this context, a new state-of-the-art approach, the fast and lightweight AutoML library (FLAML), has been shown to significantly outperform the top ranked AutoML libraries on a large set of open-source AutoML benchmarks~\citep{wang2021flaml}. We use FLAML to answer RQ1.

To answer RQ1, we investigate which algorithm FLAML has chosen as the most suitable for our dataset and then analyse feature importance measures appropriate for this algorithm. The methodology to answer RQ2 is described in detail in Section~\ref{sec:domaindiffs}.

\section{RQ1: Classifier Accuracy}
\label{sec:rq1}

In this section, we provide the answer to our first research question: How accurately can we automatically classify GitHub repositories?

\subsection{RQ1.1 What is the impact of adding different data sources on classifier performance?}

As described in Section IV-D, we used FLAML for our experiments. To investigate the impact of adding different data sources as input to the classification, we begin with only description tokens and then concatenate additional sources such as README files and labels as text data. Then we combine text data with categorical variables, i.e., contributors, programming languages, topics, contents, and licences, and finally we combine text data, categorical data, and numerical data, i.e., forks, stars, and releases. 

Table~\ref{tab:featureperformance} shows the results of these experiments. FLAML was configured to run for 1,000 seconds with 10-fold cross-validation, and due to the imbalanced nature of this dataset, we chose the macro F1 score as the validation metric~\citep{he2013imbalanced}. Although the complexity of the data increases as more features are added, the evaluation metric improves steadily as more features are added. For all the described scenarios, FLAML chose Light Gradient Boosted Machine (LGBM)\footnote{\url{https://lightgbm.readthedocs.io/en/latest/}} as the best estimator over Random forest, XGboost, Extra tree, and Linear regression, based on a 90/10 training testing split. The resulting dataset is a 1,084 dimensional vector for each of the 4,948 samples. 

\begin{table*}
\centering
\caption{Impact of adding different data sources}
\label{tab:featureperformance}
\begin{tabular}{lrrrr}
\toprule
Data Source & Precision & Recall & F1 score & Accuracy \\
\midrule
Description only & 0.58 & 0.50 & 0.52 & 0.61 \\
README only & 0.54 & 0.48 & 0.50 & 0.65 \\
Textual data only & 0.64 & 0.55 & 0.56 & 0.67 \\
Textual and categorical data & 0.64 & \textbf{0.58} & 0.60 & 0.69 \\
Textual, categorical and numerical data  & \textbf{0.70} & \textbf{0.58} & \textbf{0.61} & \textbf{0.72} \\
\bottomrule
\end{tabular}
\end{table*}

\subsection{RQ1.2: How accurately can state-of-the-art machine learning techniques automatically classify popular GitHub repositories?}

\begin{table*}
\centering
\caption{Confusion matrix achieved with FLAML, column labels are ordered in the same way as row labels}
\label{tab:confusion}
\begin{tabular}{l*6{|>{\centering\arraybackslash}m{\tabwidth}}|}
\hhline{~*6{|-}|}
     Application Software & \colorme{20} & 0 & \colorme{8} & \colorme{11} & \colorme{0} & \colorme{4} \\ \hhline{~*6{|-}|}
    Documentation & \colorme{1} & \colorme{26} & \colorme{2} & \colorme{2} & \colorme{0} & \colorme{12} \\ \hhline{~*6{|-}|}
    Non-Web Libs \& Frameworks & \colorme{5} & 4 & \colorme{108} & \colorme{10} & \colorme{2} & \colorme{14} \\ \hhline{~*6{|-}|}
    Software Tools & \colorme{3} & 3 & \colorme{10} & \colorme{69} & \colorme{1} & \colorme{10} \\ \hhline{~*6{|-}|}
    System Software & \colorme{0} & 1 & \colorme{3} & \colorme{7} & \colorme{6} & \colorme{1} \\ \hhline{~*6{|-}|}
    Web Libs \& Frameworks & \colorme{1} & 3 & \colorme{14} & \colorme{7} & \colorme{0} & \colorme{127} \\ \hhline{~*6{|-}|}
\end{tabular}
\end{table*}

From answering the previous research question, we learned that more features will likely improve the classifier's performance, but since adding features also leads to a more complex model, we explore the impact of fine-tuning the classifier in this research question. Furthermore, we use the synthetic minority oversampling technique (SMOTE) to handle the imbalance of the training data~\citep{lemaitre2017imbalanced}. SMOTE balanced the samples to 1,370 for each class, and since we now have a balanced dataset to train, we changed the validation metric to the area under the receiver operating curve (ROC-AUC) in its one-vs-rest configuration (OVR). ROC-AUC-OVR helps maximise the detection of correct samples~\citep{FAWCETT2006861}. As described in Section~\ref{sec:analysis}, we used FLAML for our experiments. Based on the results of RQ1, we also restricted the search space to only LGBM and lengthened the search time to 30,000 seconds. Table~\ref{tab:confusion} shows the initial confusion matrix achieved with FLAML, for an overall accuracy of 0.72 based on a 90/10 training testing split. The table reveals that much of the confusion was between the application domains for \textsc{Application Software} and \textsc{Software Tools}---the precision of the approach in predicting these application domains is below 0.7 and a macro averaged F1 score of 0.64. If we compare our test results with a ZeroR classifier, i.e., a classifier that always selects the most frequent class, LGBM was roughly 2.3 times more accurate (0.72 vs.~0.31) than ZeroR.   

\subsection{RQ1.3: What is the impact of merging classes of GitHub repositories?}

\begin{table}
\centering
\caption{Final classifier performance}
\label{tab:bestperformance}
\begin{tabular}{lrrr}
\toprule
Application Domain & Precision & Recall & F1 score \\
\midrule
Application \& System Software & 0.73 & 0.59 & 0.65 \\
Documentation & 0.84 & 0.74 & 0.79 \\
Non-Web Libs \& Frameworks & 0.76 & 0.77 & 0.77 \\
Software Tools & 0.72 & 0.66 & 0.68 \\
Web Libs \& Frameworks & 0.74 & 0.86 & 0.80 \\
\bottomrule
\end{tabular}
\end{table}

Taking into account that the definitions of \textsc{Application Software} and \textsc{Software Tools} are closely related (see Section~\ref{sec:background}, \textsc{Software Tools} are essentially a sub-category of \textsc{Application Software}), and that these two application domains caused confusion for our classifiers, we decided to merge those two application domains and rename them \textsc{Application \& System Software}. Note that both application domains account for a relatively small number of repositories. Anecdotally, we also observed difficulties in annotating these as separate application domains during our annotation to answer RQ1.4. We run FLAML with the same configuration as RQ1.2, and the synthetic balancing via SMOTE was still applied to the dataset.

Table~\ref{tab:bestperformance} shows that the results now improve considerably: the precision values for all application domains are above 0.7, with an overall accuracy of 0.75 and a macro averaged F1 score of 0.74. LGBM was approximately 2.4 times more accurate (0.75 vs.~0.31) than the ZeroR classifier.

\subsection{RQ1.4: Feature Importance}

\begin{table}
\centering
\caption{Feature importance}
\label{tab:featureimportance}
\begin{tabular}{llr}
\toprule
Feature Group & Example & Importance \\
\midrule
Textual features     & README file            & 94.9\% \\
Categorical features & Sub-folders/files            & 4.0\%  \\
Numerical features   & Stars                  & 1.1\%  \\
\bottomrule
\end{tabular}
\end{table}

Table~\ref{tab:featureimportance} shows the importance of the different feature groups for the classifier. According to the LGBM documentation, the importance of a feature is calculated as the number of times the feature is used in a model. 94.9\% of the prediction is explained by text data contained in README files, descriptions, and labels. Categorical data represents 4.0\% of the prediction, with sub-folders/files in the root folder and programming languages as the most important categorical data sources, while licences, contributors, and topics are not enough to explain 1\% of the total score together. Lastly, the combination of numerical data sources, stars, forks, and releases, explains the remaining 1.1\% of the feature importance score.

\subsection{RQ1.5: How accurately can state-of-the-art machine learning techniques automatically classify less popular GitHub repositories?}

Given that our classifier learns from data such as a repository's README file, we expect that it would find it challenging to classify less popular repositories, which might be poorly documented. To investigate the extent of this expected deterioration, we reached out to the authors of \cite{borges2016understanding} to help us annotate a randomly selected set of 50 less popular repositories from GitHub using the same annotation process that was used for the original dataset. The less popular repositories had a median of 169.5 stars and 41 forks each, compared to 2,866 stars and 460 forks for the popular dataset curated by Borges et al.  As expected, the performance deteriorates, with the overall accuracy dropping to 0.58. LGBM was roughly 2.2 times more accurate (0.58 vs.~0.26) than the ZeroR classifier. 

\begin{mdframed}[style=summarybox,frametitle=RQ1 Summary]
After merging two of the less common application domains that confuse human annotators and our classifier, we are able to automatically determine the application domain of a GitHub repository with a minimum precision of 0.7. The final classifier relies mostly on textual features, with categorical and numerical features playing a much smaller role. When the classifier is applied to a newly created dataset of 50 less popular repositories, the performance is lower.
\end{mdframed}

\section{RQ2: Differences between Domains}
\label{sec:domaindiffs}

To investigate to what extent differences between application domains are relevant from a software engineering perspective, we focused on three software engineering practices---automation, refactoring, and code ownership---and analysed how these differ across application domains. Since this analysis required the processing of additional data, in particular related to code changes, we took the opportunity to download a new version of the 2,000 most popular repositories on GitHub. The number 2,000 was chosen to balance effort, in particular, for downloading all corresponding commits, and size of the data. Then, we contrasted these new popular repositories with the Borges et al.~dataset and removed overlapping repositories, leaving us with 893 new unseen repositories. We downloaded all commit data from these 893 repositories on 21 March 2022. We excluded from this analysis the following four outlier repositories with more than 200,000 commits each: \textsc{chromium/chromium}, \textsc{aosp-mirror/platform\_frameworks\_base}, \textsc{llvm/llvm-project} and \textsc{Homebrew/homebrew-core}. We then automatically classified the repositories using our classifier.

Before presenting the results of the forthcoming sub-questions, we introduce a high-level view of observable differences between repositories from different application domains. We use the following nine features related to software development practices (see details in the following subsections) for each repository:

\begin{enumerate}
  \item Number of refactoring commits
  \item Number of non-refactoring commits
  \item Total number of commits
  \item Refactoring ratio
  \item Number of major contributors
  \item Number of minor contributors
  \item Total number of contributors
  \item Ownership ratio
  \item A boolean value indicating whether the repository uses automation
\end{enumerate}

With the help of T-distributed Stochastic Neighbor Embedding (t-SNE), a tool to visualize high-dimensional data, we projected all data into 2d, and then visualised the distributions for each domain in Figure~\ref{fig:density}. We can see that each of them occupies a slightly different space, with some overlaps, which is consistent with our classifier results. Visually, \textsc{Documentation} appears to have more unique characteristics in this view compared to other application domains. In the next sections, we investigate this more systematically.

\begin{figure}
  \centering
  \includegraphics[width=\linewidth]{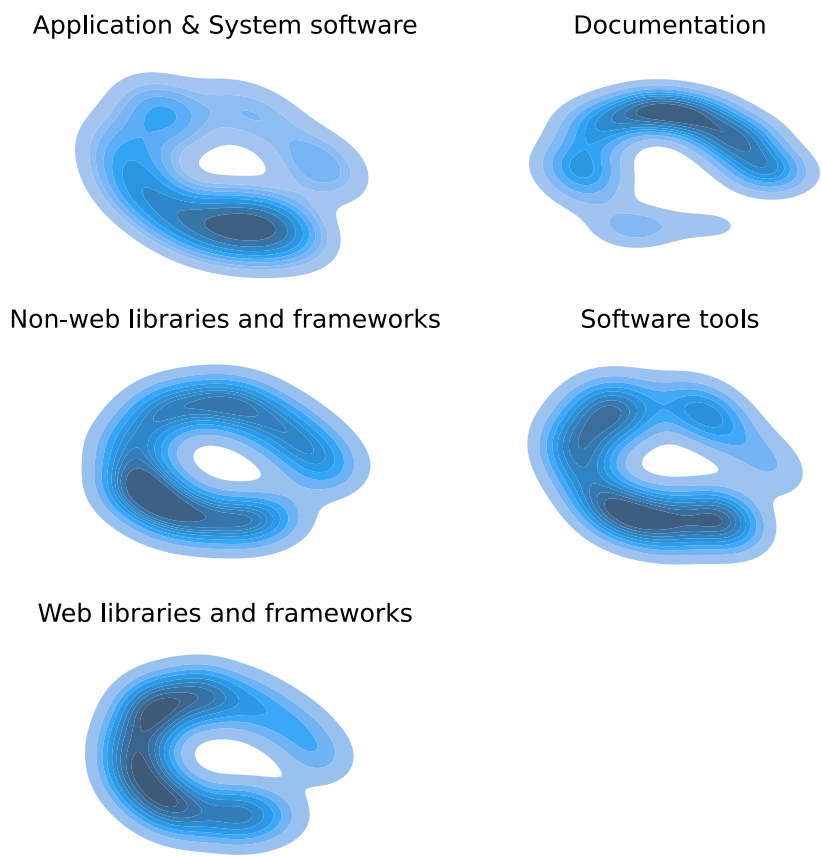}
  \caption{Density distributions across application domains}
  \label{fig:density}
\end{figure}

\subsection{RQ 2.1: How does the adoption of GitHub Actions differ between application domains?}

GitHub Actions are GitHub's answer to the increasing demand for automation in software repositories. \cite{kinsman2021software} identified that the five categories of GitHub actions that are the most frequently used are Continuous Integration, Utilities, Deployment, Publishing, and Code Quality. Because GitHub Actions standardise the use of automation by GitHub repositories, this feature affords us the opportunity to investigate differences in GitHub Actions adoption across application domains.

\begin{table}
\centering
\caption{Adoption of GitHub Actions across application domains}
\label{tab:actions}
\begin{tabular}{lrrrr}
\toprule
Application Domain                            & \multicolumn{2}{c}{Adopted} & \multicolumn{2}{c}{Not Adopted} \\
\midrule
Application \& System Software    & 55             & (70\%)            & 24                & (30\%)     \\
Documentation                     & 93              & (29\%)            & 232                & (71\%)     \\
Non-Web Libs \& Frameworks        & 97             & (56\%)            & 75               & (44\%)     \\
Software Tools                    & 104             & (68\%)            & 49                & (32\%)     \\
Web Libs \& Frameworks            & 97             & (61\%)            & 63               & (39\%)     \\
\midrule
Total                             & 446            & (50.2\%)            & 443               & (49.8\%)     \\
\bottomrule
\end{tabular}
\end{table}

Table~\ref{tab:actions} shows the results of this analysis. 50.2\% of the repositories in this dataset have adopted GitHub Actions, indicating how quickly the community has adopted this new feature. Chi-square tests of independence were performed to examine the relation between an application domain and the adoption of GitHub Actions. Following Kim~\citep{Kim2017}, we declare that an effect size $(ES) = 0.1$ is small, $ES = 0.3$ medium, and $ES = 0.5$ large~\citep{Kim2017}. The relation between these variables was significant for \textsc{Documentation} ($\chi^2 = 93.838$, $p = .000$, $phi = .325$), \textsc{Software Tools} ($\chi^2 = 22.583$, $p = .000$, $phi = .0159$), \textsc{Application \& System Software} ($\chi^2 = 12.282$, $p =.000$, $phi = .118$) and \textsc{Web Libraries and Frameworks} ($\chi^2 = 8.031$, $p = .005$, $phi = .095$). For \textsc{Non-Web Libraries and Frameworks}, the relation between these variables was not significant ($\chi^2 = 3.006$, $p = .083$, $phi = .058$).

\subsection{RQ 2.2: How does the extent of refactoring differ between application domains?}

Refactoring is the process of restructuring existing source code with the intention of improving the structure, design, and implementation of the code, but not its functionality. To analyse whether the extent of refactoring differs between application domains, we used Ratzinger's list of keywords~\citep{ratzinger2007space} to identify refactoring commits from their commit message: ``refactor'', ``restruct'', ``clean'', ``not use'', ``unus'', ``reformat'', ``import'', ``remov'', ``replac'', ``split'', ``reorg'', ``renam'', and ``move''~\citep{ratzinger2007space}. In total, 3,165,159 commits were analysed.

\begin{table}
\centering
\caption{Refactoring commits across application domains}
\label{tab:refactoring}
\begin{tabular}{lrrrr}
\toprule
Application Domain                            & \multicolumn{2}{c}{Refactoring} & \multicolumn{2}{c}{Not Refactoring} \\
\midrule
Application \& System Software    & 99k  & (14\%) & 618k  & (86\%) \\
Documentation                     & 25k   & (6\%) & 386k   & (94\%) \\
Non-Web Libs \& Frameworks        & 77k  & (15\%) & 431k  & (85\%) \\
Software Tools                    & 138k  & (13\%) & 914k  & (87\%) \\
Web Libs \& Frameworks            & 59k  & (12\%) & 419k  & (88\%) \\
\midrule
Total                             & 398k & (13\%) & 2,767k & (87\%) \\
\bottomrule
\end{tabular}
\end{table}

Table~\ref{tab:refactoring} shows the distributions of refactoring commits across application domains. Chi-square tests of independence were performed to examine the relation between an application domain and the extent of refactoring. The tests revealed that the relationship between these variables was significant, but with a small effect, for all the studied application domains: \textsc{Documentation} ($\chi^2 = 18462.478$, $p = .000$, $phi = .076$), \textsc{Software Tools} ($\chi^2 = 455.497$, $p = .000$, $phi = .012$), \textsc{Application \& System Software} ($\chi^2 = 1306.458$, $p =.000$, $phi = .002$), \textsc{Non-Web Libraries and Frameworks} ($\chi^2 = 3630.422$, $p = .000$, $phi = .034$) and \textsc{Web Libraries and Frameworks} ($\chi^2 = 20.087$, $p = .000$, $phi = .003$).

\subsection{RQ 2.3: How does the code ownership differ between application domains?}

\cite{bird2011don} developed ownership measures with the intuition that the number of times a developer works on a software component increases the developer's knowledge of that component. They defined the proportion of ownership as the ratio of the number of commits the contributor has made compared to the total number of commits. They also differentiated major contributors as developers whose ownership is at or above 5\% and minor contributors whose ownership is below 5\%. 

\begin{figure}
  \centering
  \includegraphics[width=\linewidth]{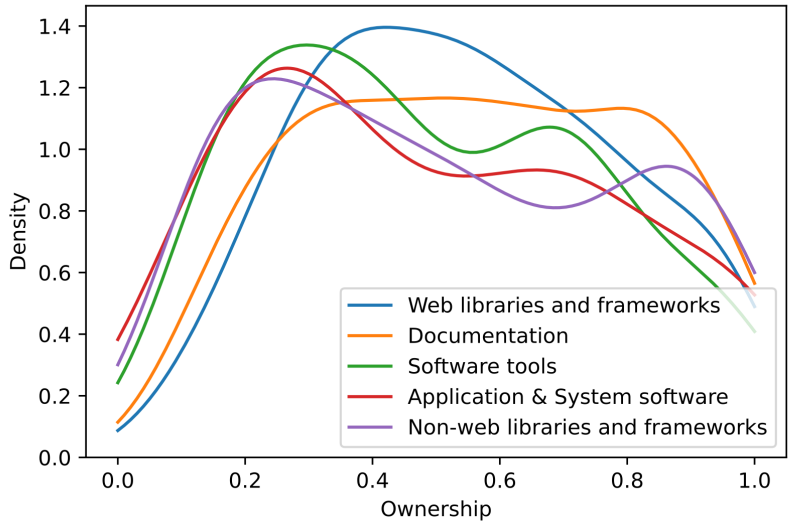}
  \caption{Ownership distributions across application domains}
  \label{fig:distribution}
\end{figure}

Figure~\ref{fig:distribution} illustrates the ownership probability density estimation (i.e., ratio of contributors identified as major contributors) for each application domain. To determine differences in these distributions, Mann-Whitney U tests were conducted on pairs of application domains. The tests revealed similar distributions for \textsc{Software Tools} (Mdn = 0.45), \textsc{Non-Web Libraries and Frameworks} (Mdn = 0.49) and \textsc{Application \& System Software} (Mdn = 0.47), but significant differences between all pairs of application domains for \textsc{Documentation} (Mdn = 0.56) and \textsc{Web Libraries and Frameworks} (Mdn = 0.54), all with small effect sizes.

\begin{mdframed}[style=summarybox,frametitle=RQ2 Summary]
Software repositories belonging to different application domains differ in their adoption of software engineering practices. Almost all of the relationships that we investigated for the practices of automation, refactoring, and code ownership revealed statistically significant differences between application domains.
\end{mdframed}

\section{Discussion}
\label{sec:discussion}

The purpose of this study is to help researchers and practitioners catalogue GitHub repositories using an automated classifier across five application domains. The advantage of this approach is that these domains would otherwise be unavailable as a GitHub feature, difficult to replicate for a novice in the domain, and highly time-consuming to re-create. As mentioned in the Introduction, many researchers have manually classified GitHub repositories using the Borges et al.~taxonomy (e.g.,~\cite{bi2020empirical, bi2021first, hayashi2019impacts, rehman2022newcomer, hata2022github, de2022impacts, de2020splicing, tamburri2019discovering, coelho2017modern})---our classifiers aim to eliminate the need for such manual work.

Strengths of this high-level application domain taxonomy include the ability to combine it with other filtering methods, such as topic tags, to help narrow down the search of particular repositories. On the other hand, taxonomy and automated classifier enable discovery of common characteristics of projects in a specific domain, leading to generalisable results.

From our results of RQ1, we learned that more data sources as features were beneficial in improving our classifier performance and that it was hard to find features to separate \textsc{Application Software} from \textsc{Software Tools}. We observe that separating these two categories was not a trivial task, even for a domain expert, so we ended up merging them. Textual data, i.e., descriptions and README files, are the most important drivers to catalogue a repository. Maintainers looking for contributors should make sure that their README file is detailed and up-to-date, since it is the main source for classifiers to categorise a repository. README files are also the starting point for developers when encountering a new repository~\citep{prana2019categorizing}. We hypothesise that poor descriptions might influence lower results for less popular repositories, but further research is needed to corroborate this.

RQ2 results show how our classifier should be used rather than manually annotating the repositories. The cataloguing time took only a fraction of the time of doing it manually and the initial data was easily complemented with more data. The findings revealed some overlap between application domains, with a clear distinction for Documentation, which, despite having a significant relationship between the adoption of GitHub domains and the extent of refactoring commits, has the lowest rates of adoption of GitHub actions (Table~\ref{tab:actions}) and of extent of refactoring (Table~\ref{tab:refactoring}), as well as a dissimilar distribution of code ownership. As such, the reliability of these findings is based on their ability to be easily reproduced by anyone, fulfilling our earlier promise of aiding in the achievement of generalisable results.

\section{Threats to Validity}
\label{sec:threats}

The dynamic nature of GitHub is a threat to the validity of this work. What was considered a popular repository by \cite{borges2016understanding} may not be a popular repository today, and that definition influenced what was considered less popular in order to test the accuracy of this model. This dynamic nature also influenced why some repositories could not be downloaded to answer RQ2 and whether their presence might have influenced the results.

The application domains used in this work are based on a manually annotated dataset with all the imperfections that might be present in terms of bias and errors. However, every effort has been made to keep those criteria in place throughout the work in order to make it as consistent as possible. The descriptions of what should be considered one or another application domain leave some room for interpretation, despite the fact that manually annotating 5,000 repositories is a substantial effort and there are few opportunities to use such a valuable resource to create a classifier as this one. This threat was mitigated by the authors of this classification who debated difficult decisions and the involvement of Borges et al.~in the annotation of less popular repositories to ensure consistency across the annotations.

Further threats to validity exist in the fact that keywords to find refactoring commits can be imperfect or incomplete, despite us relying on a previously published set of keywords for the identification of such commits. In addition, contributors can make their commits under more than one user name, affecting code ownership metrics, and part of a repository's history may not have been completely migrated to what is contained in GitHub, affecting results. All of these are difficult to control, and interpretations of results must take these into account. Finally, to answer RQ2, four repositories were removed from the dataset. Due to the large number of commits compared to the other repositories, these four repositories \textsc{chromium/chromium}, \textsc{llvm/llvm-project} and \textsc{Homebrew/homebrew-core} may have affected the extent of the refactoring results for the \textsc{Software tools} application domain, as well as \textsc{aosp-mirror/platform\_frameworks\_base} may have impacted the results for the \textsc{Non-web libraries and frameworks} application domain. 

\section{Related Work}
\label{sec:related}

To our knowledge, our classifier is the first automated tool to predict the application domain of a GitHub repository. It follows in the long line of work on classifying other aspects of GitHub repositories, such as the content of README files or a repository's topic tags.

In 2014, when there were around 10 million repositories on GitHub, \cite{kalliamvakou2014promises} predicted the importance of this platform for researchers, as well as the difficulties in distinguishing real software repositories from repositories for personal use. Although features such as topic tags are implemented on GitHub, most repositories do not use them~\citep{zhang2019higitclass}. Therefore, researchers have been investigating approaches to assist GitHub users in finding repositories of their interest and repository owners in improving their documentation. 

Following this reasoning, \cite{sharma2017cataloging} extracted descriptive segments of text from 10,000 README files in GitHub repositories. They pre-processed those descriptive segments with standard natural language processing techniques, i.e., tokenisation, stop word removal, and stemming, and evaluated them using the Latent Dirichlet Allocation Genetic Algorithm (LDA-GA), to identify a set of topics together with the repositories belonging to those topics. This is a promising approach with a resulting F1 score of 0.7 and can be helpful in predicting missing topic data for some repositories. However, the process involves manual analysis, which makes it difficult to operate on new data.

\cite{ma2018automatic} proposed an automated solution based on machine learning techniques to identify different types of software artefacts, such as requirements documents, system elements, verifications, or tasks. On a test dataset, they were able to automatically categorise documentation-related software artefacts with an average precision of 0.76 and a recall of 0.75 by combining several machine learning techniques through ensemble learning.

\cite{zhang2019higitclass} proposed a hierarchical classification that considers multimodal signals in a repository, i.e., user names, descriptions, tags, and README files. They proposed a framework composed of three key modules: a graph containing those multimodal signals defined as ``heterogeneous information network'', keyword enrichment, and pseudo-document generation. Finally, the framework was tested on two GitHub datasets containing separated data from the machine learning and bioinformatics research communities against six hierarchical classification algorithms, achieving outperforming results.

\cite{prana2019categorizing} conducted a qualitative study of README files in GitHub repositories and developed a classifier to automatically identify eight categories of README sections and achieved an F1 score of 0.746. They find that discussion about ``What'' and ``How'' of a repository is quite common, while information about purpose and status is limited or missing. Then, they labelled sections of README files using badges and tested the classification outcome with the opinions of twenty software professionals, who in the majority perceived ease of information discovery. Finally, they provide recommendations to repository owners to improve their documentation quality.

\cite{di2020multinomial} trained a multinomial Naïve Bayes network with TF-IDF vectors computed from README files to predict a list of topics. Then, they combine those results with discovered programming languages using the GuessLang tool to deliver a resulting list of recommended topics. According to the analysis of their results, the model performs well with a low number of topics to predict, i.e., one or two. However, their results decrease notably as the number of recommended topics increases.

Complementing these papers, our work focuses on the automated classification of repositories according to their application domain, based on the domains established by the work of \cite{borges2016understanding}.

\section{Conclusion and Future Work}
\label{sec:conclusion}

Past research has found that the application domain of a GitHub repository is an important piece of information, for example, to predict the popularity of a repository~\citep{borges2016understanding} or as a parameter when assessing the quality of the code contained in the repository~\citep{capiluppi2019relevance}. Yet, determining the application domain of a GitHub repository is a non-trivial task.

In this work, we use an existing dataset of 5,000 GitHub repositories as a starting point for building a classifier which can automate this task. Each repository in the dataset has been manually annotated with its application domain, and we augment this dataset with textual, categorical, and numerical features that capture the characteristics of each of the repositories.

We show that our automated classifier is able to assign popular repositories to their application domain with at least 70\% precision for each of the five domains, and we explore the impact of various design choices in the creation of the classifier in detail. We then apply the classifier to a new dataset of repositories to investigate to what extent repositories associated with different application domains differ from a software engineering perspective. In particular, we investigate the adoption of software engineering practices, such as automation and refactoring, across the application domains, uncovering significant differences.

\vspace{0.4cm}
\noindent
\textbf{Future Work. }
To further improve the performance of the classifier, future research should consider additional options to address the issue of incomplete documentation in some repositories. Although topic tags have a high predictive capacity, they are not used in most repositories studied. These techniques are also applicable when dealing with less popular repositories, such as those discussed in RQ 1.4. For a repository with little documentation, our classifier might be able to help users understand the repository by identifying its application domain.

Promising results can be obtained by combining classifiers and datasets; for example, one classifier could predict application domains using only text data, while the other classifier would use the remaining data sources, perhaps one for categorical data and another for numerical data. This proposed method can be helpful in diminishing the complexity of each dataset-classifier pair and will likely further improve the results.

Meanwhile, our classifier enables other applications that need to quickly and automatically sift through the large number of repositories on GitHub. We found that the adoption of software engineering practices differs depending on the application domain---this has implications for practitioners and educators alike. For example, one interesting question to explore would be whether the observed differences imply that students need to be taught differently depending on their intended application domain or whether practitioners working in certain application domains should be encouraged to adopt the practices of others. Our classifier enables the collection of a large number of repositories to drive such work.

\bibliographystyle{spbasic}

\end{sloppy}
\end{document}